# Understanding Tool Synthesis Behavior and Safe Finite State Machine Design

Timothy McDonley
Cyber Trust and Analytics
Battelle Memorial Institute
Ohio, USA
mcdonley@battelle.org

Katie Liszewski
Cyber Trust and Analytics
Battelle Memorial Institute
Ohio, USA
liszweski@battelle.org

*Abstract*— High-reliability design requires understanding synthesis tool behavior and best practices. Detection and protection against illegal states and transitions is important for critical Finite State Machines (FSMs) within high reliability applications. Single Event Upsets (SEUs) probability is increasing with decreasing circuit dimensions and voltage [1]. SEU handling must be analyzed post optimization to ensure designed protections are still functional. In this work the default behavior of three synthesis tools interacting with high reliability FSMs is discussed. Post-synthesis netlists of test FSMs are analyzed for optimization induced changes that affect reliability during a SEU. Best practices are proposed to curtail aggressive optimizers.

*Keywords— Finite State Machine, Single Event Upset, High-Reliability, Synthesis*

## I. INTRODUCTION

High-reliability designs are critical in a wide range of applications and utilize both application specific integrated circuit (ASIC) and field programmable gate array (FPGA) workflows. Whether it be industries like automotive or medical, industrial safety systems, or remote operation, understanding the behavior of the design outside of normal operation and the potential for down time during a reset is paramount. A single event upset (SEU), where a bit value in a register is flipped unexpectedly, is exceedingly rare but can occur in either an ASIC or FPGA and, if poorly handled may require a system restart to recover. FPGA manufactures [2] [1] provide best practices for mitigation and metrics of SEU rate in their hardware. Detection, mitigation, and correction are a critical design focus in mission critical systems development that involve FPGAs [3] or ASICs [4].

After a SEU, a finite state machine (FSM) that does not have proper protection could enter an illegal state causing it to lock up, leaving the system inoperable for an indeterminant amount of time. The device could also enter a non-recoverable transition loop that requires a reboot of the entire system. A safe FSM seeks to address these issues through a combination of detection, prevention, and tolerance. A safe FSM might use detection of illegal states to fail in a known and acceptable way. Certain illegal states may be tolerable if the design can recover in the next clock cycle and continue with an acceptable transition.

During synthesis, the FSMs in a design can be re-encoded and altered in many ways to improve optimization or to mitigate poor coding practices. Understanding the synthesis behavior of the chosen tool is just as important as the coding practices used in the design. Many tools favor optimization, a common requirement, over SEU tolerance, an uncommon requirement, in re-encoding during synthesis. A lack of understanding of default synthesis behavior and the directives available can cause a designer to inadvertently remove SEU mitigations during optimization.

## II. METHODOLOGY

The ISCAS-89 S298 benchmark is chosen to illustrate SEU effects as it is small enough for easy visualization of the state space. This design benchmark mimics a traffic light controller and depends heavily on a modulo 10 counter that outputs 4-bit unsigned binary. It is shown that SEUs on the traffic light controller can cause dangerous traffic patterns that cannot be recovered from until the device is reset manually. The effects of synthesis using standard settings for both an ASIC and FPGA workflow on this FSM are detailed. The design tools used in this paper were Xilinx Vivado (FPGA) [5], Yosys (open-source synthesis) [6], and Cadence Genus (ASIC) [7]. The authors wish to emphasize these results do not indicate failure of any one design tool, and rather simply a need for community awareness of tool behavior when designing against stringent requirements.

The FSM is extracted from the ISCAS-89 S298 netlist using the netELA [8] tool. The netELA tool provides an automated means for identifying potential state registers, locating all relevant feedback and control logic, and extracting a target FSM, even in a flattened post-synthesis netlist. The extraction produces an SMT based model of the FSM. This model, which is restricted to FSM standard cells, allows for efficient exploration of the FSM state space, including

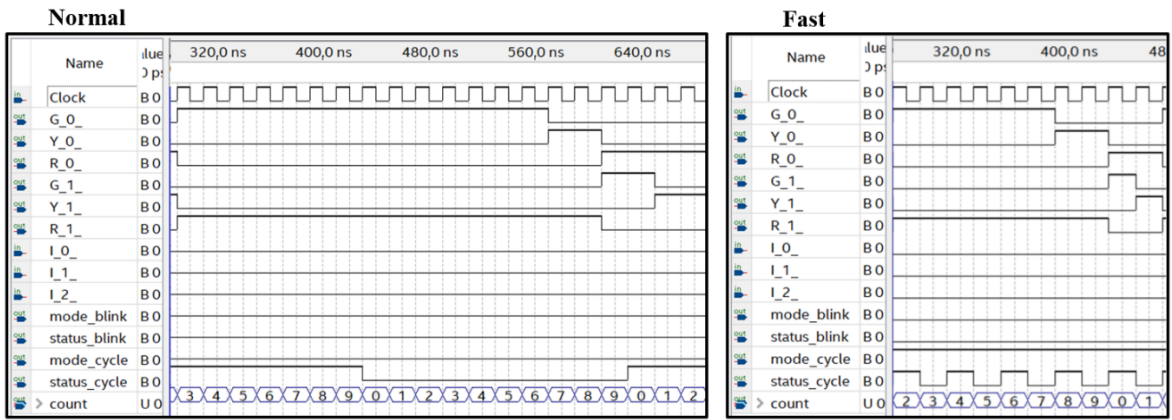

Figure 1: ISCAS-89 S298 One cycle of normal and fast operation modes

satisfiability across multiple clock cycles and automated test bench generation.

While traditional simulation and verification tools do offer methods to corrupt an FSM and explore behavior after an SEU, they represent a considerable investment of time and effort to use [1]. The FSM model extracted by netELA offers ways of interacting with the FSM not possible with traditional simulation techniques through SMT solver techniques. For instance, a designer may need to confirm that there is no possible path to a specific state, including through a SEU. The netELA SMT solver-based model of the FSM provides a means for modeling this assertion over multiple clock cycles. The satisfiability of the assertion can be checked without having to simulate the FSM on every possible state transition. The FSM model will either return that the state is unreachable or provide an example set of state transitions and control inputs to reach the specified state (not necessarily a unique solution). In this way the designer is essentially running the state transitions in reverse, starting at the problem state, determining if the issue exists and if so, working backwards into one of many known states. Conversely, traditional simulation would need to create a test bench with full coverage, including modeling SUEs, to definitively know that a state is unreachable. This is of great importance as an FSM can become so large that simulation of the entire state space becomes computationally untenable.

State Transition Graphs (STGs) are presented in this paper using a graph visualization tool [9]. An FSM module in the design is reencoded with multiple encodings and the full design synthesized in Vivado, Yosys, and Genus. The FSM is then extracted from the post-synthesis netlist and the effects of optimization are explored.

### III. POST SYNTHESIS NETLIST FSM ANALYSIS

The ISCAS-89 S298 netlist describes a traffic light controller that directs two directions of traffic, named primary and secondary. For the primary lane, the sequence is 14 clock cycles of green, 2 clock cycles of yellow, and 4 clock cycles of red. For the secondary lane, the sequence is 16 clock cycles of red, 2 clock cycles of green, and 2 clock cycles of yellow. This is shown in Figure 1 (left). The device also includes a faster cycle mode, shown in Figure 1 (right). For the primary lane, the fast sequence is 6 clock cycles of green, 2 clock cycles of yellow, and 2 clock cycles of red. For the secondary lane, the fast sequence is 8 clock cycles of red, 1 clock cycles of green, and 1 clock cycles of yellow. The device also includes an emergency 'blink' mode which is not explored in this paper.

The timing of the cycles is done with the *Counter* module, a modulo 10 counter, and the *status_cycle* register. The *status_cycle* register keeps track of the first half and second half of the 20 stage cycle in conjunction with the counter's 10 states. In the fast mode operation, which is only 10 cycles, the *status_cycle* register is logic high on even counter states and low on odd states.

Extraction of the FSM from the original ISCAS-89 S298 netlist is done automatically with the netELA tool using a connectivity-based force-directed clustering algorithm that forms groups using a Gaussian Mixture model [8]. Unsurprisingly, the four bits of the counter are grouped together with the *status_cycle* register. The feedback logic from the group is extracted and used to generate the STG shown in Figure 2.

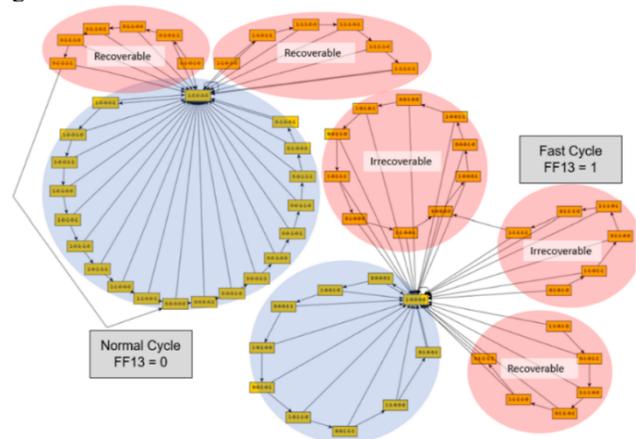

Figure 2: State Transition Graph (STG) for S298 Counter module and *status_cycle*.

The state transition loops highlighted in blue are the legal states, as defined by the design specifications, and result in proper device operation. The state transition loops highlighted in red show illegal states, in this case counter values between 10-15. These transitions drive the other modules in the design and cause unexpected behavior that is explored below. Recoverable loops are illegal transitions that lead back to normal device operation after a number of clock cycles.

Irrecoverable loops are illegal transitions that do not return to normal device operation without a manual reset signal. Note that for this device there is no deadlock; a state that can be reached, but not exited from. The irrecoverable loops are given by the *status_cycle* register getting out of sync from the counter and having the wrong even/odd relationship.

The STG and illegal states can be used to understand how SEUs might affect the device as a whole. For brevity and clarity, this paper choses a scenario to discuss, shown in Figure 3. In this case, a SEU puts the device into an irrecoverable loop of illegal states. The state transitions and associated control values are used to generate a test bench to drive the design and determine effects. When the device is in the fast cycle, a SEU can move the state from 2 to 10. The STG shows that the counter will continue to 15 before returning to 0. Unfortunately, it will not recover because the *status_cycle* register will be out of sync.

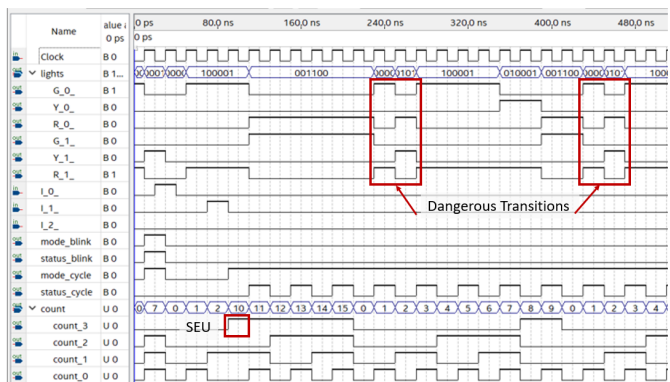

Figure 3: SEU puts S298's counter module into an irrecoverable loop. The loop of illegal state transitions causes dangerous switching between Green and Red traffic signals on both lanes.

*A. Gray Code Encoding*

In exploring synthesis tools' handling of FSMs, the *Counter* module is replaced with a 4-bit Gray code encoded FSM with transitions for states 0-9 defined in a case statement. The FSM includes a count up or down option to provide some complexity but is always held at count up for testing. The default case is a transition back to state 0. A second case statement decodes the state back into unsigned decimal. However, this case statement does not provide a default case. The design was then synthesized using Vivado, Yosys, and Genus. All settings were left at their defaults. The post synthesis netlists were verified using the ISCAS-89 S298 testbench and then the *Counter* module was extracted using the netELA tool as described above.

*1) Vivado*

The optimizer chose to automatically reencode the counter module with a one-hot encoding. The registers and the latches for the decoded count values include clock-enable lines. However, the signal logic for clock-enable only disables state transitions if the state is zero-hot. The result is shown in the STG of the extracted counter, Figure 4. While the provided default case was a transition to state zero, the STG shows that a wide array of illegal states and transitions are possible in the reencoded counter. The output of the module is not protected. Due to this, two-hot states are decoded to values between 0-15 based on the one-hot decoding logic. In this case a SEU can lead the counter into an irrecoverable, non-sequential pattern of output values that includes numbers between 0-15. In general, a switch to one-hot leads to simplified detection of illegal states. However, because protections on the outputs are not implemented for two-hot states this FSM has a very high likelihood of driving the device improperly during a SEU as the ratio of illegal to legal states is also high.

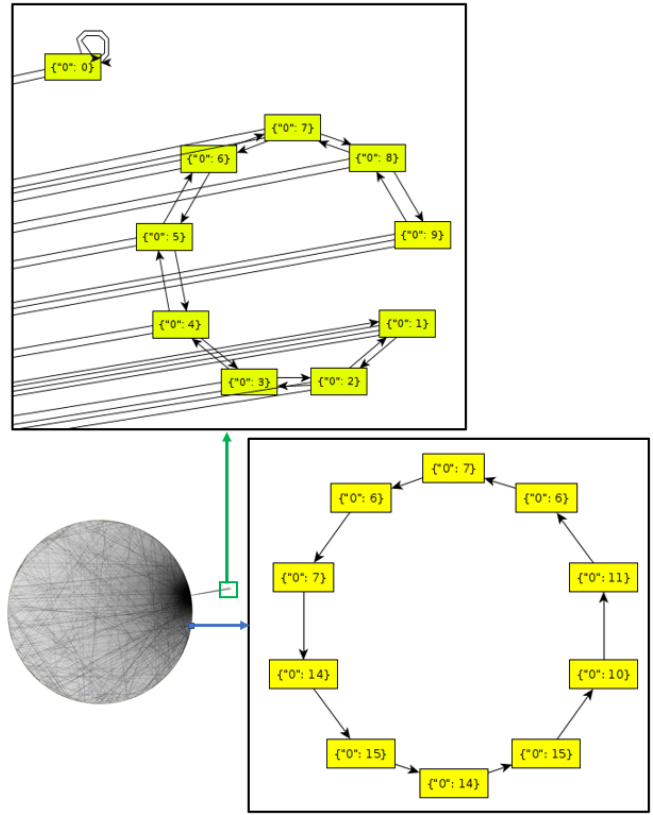

Figure 4: STG of post synthesis extracted counter reencoded in one-hot for Vivado. Main transition loop (top). Illegal states/transition(left). One example of irrecoverable loop from illegal state 10(right).

*2) Yosys*

This tool's synthesis also reencoded the counter to one-hot. In this case the outputs of the module are latched with the enable line only preventing the outputs from following a zero-hot as well. The STG, shown in Figure 5, is similar to Vivado in terms of number of transitions and illegal states. While the actual logic is different, the method for decoding from the state register's one-hot to unsigned binary is the same. In this case the Vivado and Yosys post synthesis designs share the same irrecoverable loop.

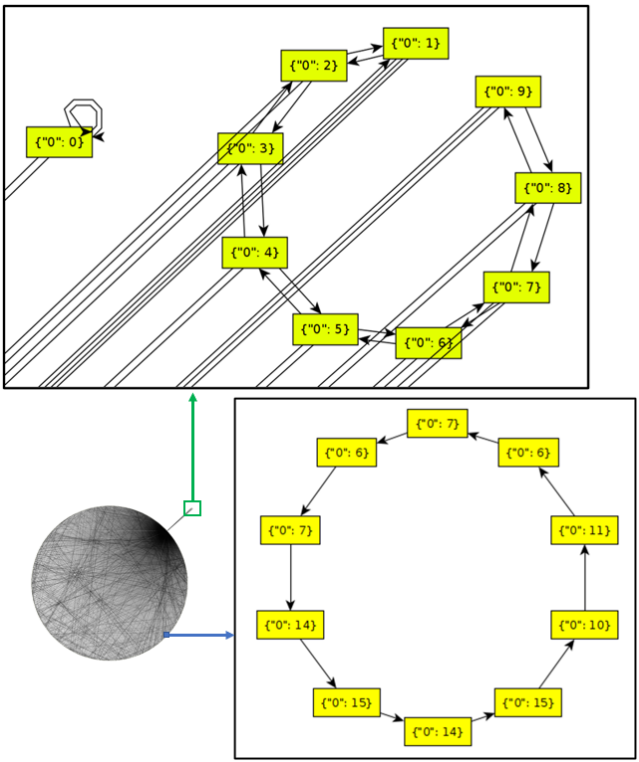

Figure 5: STG of post synthesis extracted counter reencoded in one-hot for Yosys. Main transition loop (top). Illegal states/transition(left). Example of irrecoverable loop from illegal state 10(right).

*3) Genus*

This tool's synthesis did not reencode the FSM during optimization. The extracted counter shows that optimizations did not add any additional state transitions, seen in Figure 6. Output of combinational decoding logic from the state register is latched. A common enable line on the latches prevents the output of Gray code states for 10-15. In this case, the unreachable states would not propagate out of the counter. The latch line would not be enabled and the counter would hold its previous value. Then the counter would transition to 0.

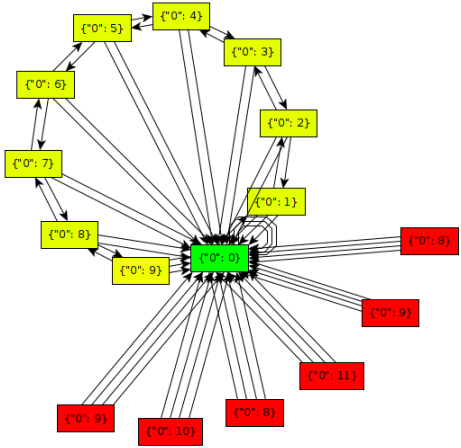

Figure 6 STG of post synthesis extracted counter for Genus. Illegal states (shown in red) are unreachable and in the event of a SEU transition to 0 as defined in the default case. The decoded value of the illegal Gray states would not propagate out of the counter and the previous valid value would be held.

## B. Binary Encoding

Here the counter module is replaced with a simple binary encoded counter with a case statement defining the transitions from 0-9 and a default case of 0. Again, the design was synthesized using Vivado, Yosys, and Genus for comparison. In this case the Vivado and Yosys do not use similar optimization strategies as they did in III.A.

*1) Vivado*

The post synthesis netlist shows that Vivado again reencoded the counter module to one-hot. In this instance there are no latches or enable lines used, even for zero-hot. The STG of the extracted post synthesis counter shows a situation largely the same is in III.A.

*2) Yosys*

Here, we see a major difference between the post synthesis FSMs generated by the Vivado and Yosys. For Yosys, synthesis leaves the encoding alone and the case statement follows the RTL. The counter will recover from any illegal state by transitioning to the default case of zero. No latches or enable lines are used, so the illegal state will propagate out of the counter module and influence the device.

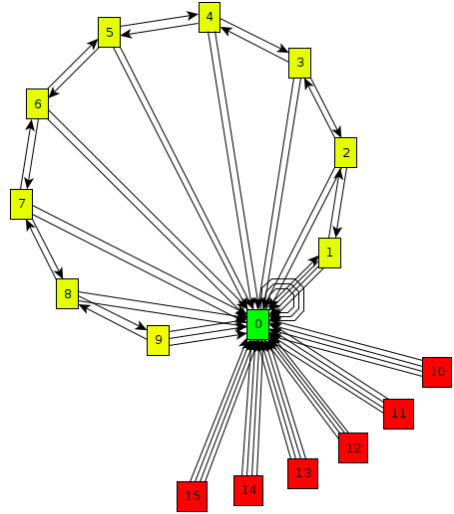

Figure 7: STG for post synthesis extracted the binary encoded counter module for Yosys.

*3) Genus*

This tool's synthesis did not reencode the FSM during optimization. The extracted counter shows that optimizations did not add any additional state transitions, seen in Figure 8. When unexpected states are encountered, the default case provided in the RTL is followed and the counter transition to 0 at the next clock edge. In this instance the STG post-synthesis matches the STG of the provided RTL.

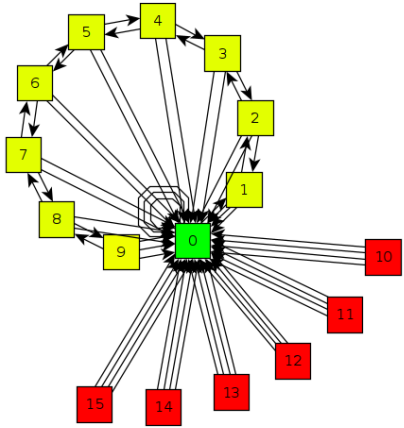

Figure 8 STG of post synthesis extracted counter for Genus. Illegal states are unreachable and in the event of a SEU transition back to 0 as defined in the default case.

### C. Best Practices

As discussed in the explorations above, different tools prioritize aspects of FSMs differently during synthesis. Each tool provides methods to assert control over the synthesis of the FSM and some of the tools have various levels of means to ensure the output of a safe FSM. This is typically done through the combination of synthesis settings, directives, and labels. The case shown in III.A for Vivado can be dealt with properly, provided the user is aware of the synthesis behavior and how to rectify it. A label was added to the Gray code counter providing the Vivado with a directive to not reencode the FSM. The case statement was given a label to tell Vivado that the cases are fully defined. The FSM extracted from the post synthesis netlist shows a STG that is as defined in the RTL and better able to handle SEUs. The illegal states presented by the Gray code for 10-15 transition to state 0 on the next clock cycle. The output of the counter does not allow values over 9 to propagate out of the module.

Combining coding practices targeted toward detection and/or correction of SEUs [10] and awareness of proper synthesis settings further improves reliability and safety. For the Traffic Light Controller example and similar safety critical systems preventing illegal states and transitions, as demonstrated in Figure 9, is not generally sufficient. Correcting the illegal states by transitioning back to 0 could itself produce dangerous behavior depending on when the reset occurs. Furthermore, abruptly transitioning to a legal state through a SEU could be equally undesirable. In this case an encoding with a Hamming distance of three between states as suggested in [10] results in safe correction. The encoding may need to be done by hand depending on choice of synthesis tool. In this case Vivado provides some support for automatically re-encoding target FSMs with different protection or detection schemes, including Hamming 3. Yosys and Genus however do not provide any detection or protection focused re-encoding schemes. In the case of hand coding, a designer must choose optimization settings that preserve this encoding.

Figure 10 shows the STG of the post-synthesis extracted Counter module from Vivado with the Hamming 3 re-encoding setting. Red nodes have no transition into that state. The red nodes in the main ring represent all possible SEUs and show they properly retain the correct current state value once decoded and transition properly to the next state. One example of a SEU is shown in the figure's break-out box. The valence semi-circle represents multiple even upsets which cannot be corrected by a Hamming 3 encoding. Vivado transitions the device back to the default state in this case. Figure 11 shows the result of the combined synthesis settings and Hamming 3 encoding discussed on the SEU shown in Figure 10. The device properly retains its current state, transitions to the correct next state, and continues to provide valid behavior. Although this paper focused on the *Counter* module, other registers in the device would also need to be hardened. The relationship between the counter and the *status_cycle* register in the *Status* module is an example of such.

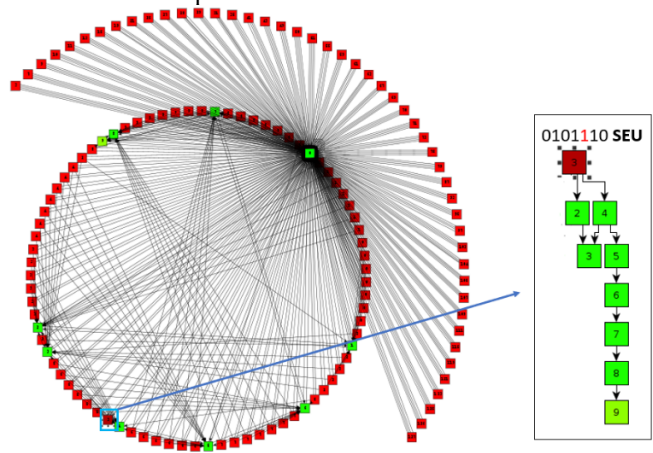

*Figure 10: Extracted Counter STG from post Vivado synthesis netlist with Hamming 3 re-encoding.*

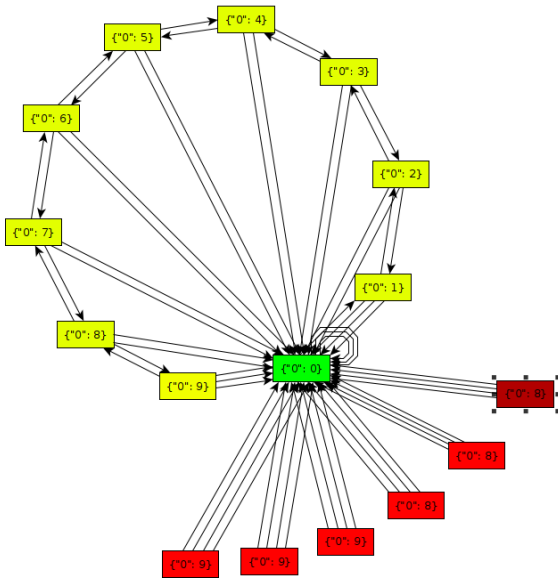

Figure 9: STG for post synthesis extracted Gray code encoded counter module for Vivado with case and FSM labels.

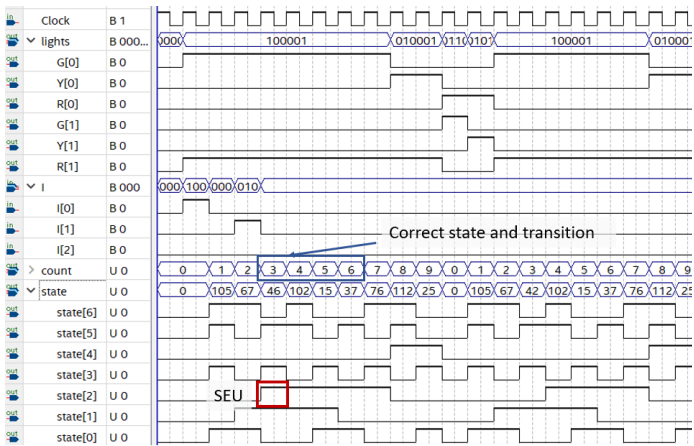

*Figure 11: SEU is properly handled. Counter retains correct state and properly transition. No unwanted traffic signals.*

Best practices for safe FSM start with good coding practices that include using appropriate detection and correction schemes [10], but they do not end there. The explorations above provide an understanding that the burden of producing a safe FSM post synthesis should be assumed to be placed on the user. As such, the user has a responsibility to understand the synthesis behavior of the tool they are using given the settings, directives and labels provided.

## IV. Conclusion

FSM design and encoding are key components of high reliability design, but best practices should not end at the design phase. Knowledge of a chosen synthesis tool's default behavior, the options available to the user, and how these options affect the synthesized design are critical. During the synthesis process an FSM's encoding and protections can be altered or replaced depending on the settings, directives, and labels used or not used. Many tools favor optimization over a safe FSM in a default re-encoding during synthesis. A lack of understanding of the synthesis behavior and the directives available can lead to even a well-designed device being vulnerable to SEUs.

This paper has explored the default behaviors of three different synthesis tools to provide a clearer understanding of how synthesis affects an FSM with respect to SEUs. With a better knowledge of optimization in their tool flow a designer should appreciate the responsibility thrust upon them to correctly manage the synthesis process.